\documentclass[journal,11pt,onecolumn,draftcls]{IEEEtran}
\usepackage[left=1.25in, right=1.25in, bottom=1in, top=1in]{geometry}

\usepackage{cite}
\usepackage{url}
\usepackage{todonotes}
\usepackage{amsmath,amssymb,amsfonts}
\usepackage{algorithmic}
\usepackage{graphicx}
\usepackage{textcomp}
\usepackage{xcolor}
\usepackage{tabularx}
\usepackage{booktabs}
\usepackage{multirow}
\usepackage{collcell}
\usepackage{hhline}
\usepackage{tikz}
\usepackage[normalem]{ulem}
\usepackage{pgfplots,etoolbox}
\usepackage{pgfplotstable}
\usepackage{pgffor}
\usepackage{ifthen}
\usepackage{float}

\pgfplotsset{compat=1.18}
\graphicspath{ {./Figures/} }

\begin{document}

\title{Information Forensics and Security:\\ A quarter-century-long journey}

\author{M. Barni, P. Campisi, E.~J. Delp, G. Do{\"e}rr, J. Fridrich, N. Memon, \\ F. P{\'e}rez-Gonz{\'a}lez, A. Rocha, L. Verdoliva, and M. Wu}

\date{March 2023}

\maketitle

\section*{Abstract}
Information Forensics and Security (IFS) is an active R\&D area whose goal is to ensure that people use devices, data, and intellectual properties for authorized purposes and to facilitate the gathering of solid evidence to hold perpetrators accountable. For over a quarter century since the 1990s, the IFS research area has grown tremendously to address the societal needs of the digital information era. The IEEE Signal Processing Society (SPS) has emerged as an important hub and leader in this area, and the article below celebrates some landmark technical contributions. In particular, we highlight the major technological advances on some selected focus areas in the field developed in the last 25 years from the research community and present future trends.
\section{Introduction}

The rapid digitization of society over the last decades has fundamentally disrupted how we interact with media content. How can we trust recorded images/videos/speeches that can be easily manipulated with a piece of software? How can we safeguard the value of copyrighted digital assets when they can be easily cloned without degradation? How can we preserve our privacy when ubiquitous capturing devices that jeopardize our anonymity are present everywhere? How our identity is verified or identified in a group of people has also significantly changed. Biometric identifiers, used at the beginning of the twentieth century for criminal investigation and law enforcement purposes, are now routinely employed as a means to automatically recognize people for a much wider range of applications, ranging from banking to electronic documents to Automatic Border Control systems, to consumer electronics. 

While the issues related to the protection of the media content and the security of biometric-based systems can be partly addressed using cryptography-based technologies, complementary signal-processing techniques were needed to address them fully. It is those technical challenges that gave birth to the IFS R\&D community. Primarily driven by the need for copyright protection solutions at an early age, IFS contributions were published in various venues and journals that were not dedicated to this area. Although some dedicated conferences (SPIE/IST Security, Steganography and Watermarking of Multimedia Contents; ACM Multimedia and Security; Information Hiding Workshop) emerged, this nascent community lacked a well-identified forum where researchers, engineers, and practitioners could exchange about the latest advances in this area which is multidisciplinary by nature. A call for contributions of the IEEE Transactions on Signal Processing in 2003 attracted enthusiastic responses to fill three special issues on Secure Media. It was time to create a platform to advance the research and technology development of signal processing-related security and forensic issues. 

To foster broader community building and strive for a bigger and lasting impact, a collective effort by a group of volunteer leaders of the IEEE SPS charted a roadmap in 2004 for creating the IEEE Transactions on Information Forensics and Security (IEEE T-IFS) and a corresponding IFS Technical Committee, both of which were launched in 2006. It was written in the proposal to the IEEE that the new journal would aim at examining IFS issues and applications “both through emerging network security architectures and through complementary methods including, but not limited to:  biometrics, multimedia security, audio-visual surveillance systems, authentication and control mechanisms, and other means.” A few years later, in 2009, the first edition of the IEEE Workshop on Information Forensics and Security (WIFS) was held in London, UK.

The IFS community has established a strong presence in the IEEE SPS and is attracting submissions from a variety of domains. In view of the page budget allocated to this retrospective article, rather than surveying, exhaustively but briefly, each individual IFS area, we opted for a more focused review of selected domains that experienced major breakthoughs over the last 25 years and that are expected to be more aligned with the interests of the Signal Processing Magazine readership. While this authoring choice does imply that some IFS areas will not be covered, it shall not be taken as any indication that some IFS contributions are more welcome than others. We hope that the following few pages will give the readers a flavour of what happened in this field, as well as the specifics of the mindset underpinning this research area.

\section{Digital Watermarking}
\label{sec:watermarking}

In the late 90s, MP3 song exchanges on peer-to-peer file-sharing networks and DVD ripping increased piracy concerns. In this emerging digital, interconnected world, generational copies became perfect clones that could be efficiently distributed worldwide without being burdened by the shipping logistics of the old analog world. Digital watermarking was introduced in this context to complement traditional cryptography-based solutions and provide a second line of defense. The essence of this technology is to introduce imperceptible changes in media content --~should it be audio, images, video, text, or other~-- to transmit information that can later be recovered robustly, even if the watermarked content has been modified \cite{wm:WPD99}.

This new research area rapidly attracted contributions from related domains: perceptual modeling, digital communications, audio/video coding, pattern recognition, etc. Early watermarking methods used very simple rules, e.g., least significant bit replacement, thereby providing almost no robustness to attacks. Significant progress was made when the IFS research community realized that the retrieval of the embedded watermark could be framed as a digital communications problem. A seminal watermarking contribution, coined as spread-spectrum watermarking~\cite{wm:CKL+97} leverages a military communications model well known for its resilience to jamming. The underlying principle is to spread each watermark bit across many dimensions of the host media content to achieve robustness; for a given bit $b \in \{\pm1\}$ to be embedded and an input (host) $n$-dimensional vector $\bf x$ additive spread-spectrum outputs a watermarked vector $\bf y$ such that 
\begin{equation}
\label{eq:add_ss}
    {\bf y}={\bf x}+ b {\bf w}
\end{equation}
where $\bf w$ is an $n$-dimensional carrier secret to adversaries. Spreading is achieved when $n \gg 1$. Due to intentional or inadvertent attacks (e.g., content transcoding), a legitimate decoder (that knows the carrier $\bf w$) gets access only to a distorted version of $\bf y$ from which it must extract the embedded bit $b$ with the highest possible reliability.         
By further exploiting connections with statistical detection and coding, it has been possible to derive optimal ways to extract the embedded watermark information for various hosts and increase robustness using channel coding. It should be kept in mind, however, that watermarking deviates from standard communications theory in that:
\begin{enumerate}
    \item the watermark embedding process must remain imperceptible, so standard power constraints (i.e. $||{\bf w}||^2/n\ll||{\bf x}||^2/n$) are not sufficient, and
    \item the wide range of attacks that the watermark is expected to be resilient to exceeds the typical channel distortions and jammers.
\end{enumerate}
Spread-spectrum watermarking fundamentally assimilates the host media content $\bf x$ to interference on the underlying low-powered watermark transmission $b \bf w$, which can, at best, be modeled statistically.

A breakthrough came about by accounting for the fact that the media host is fully known at the time of watermark embedding (but not watermark detection) and that an alternative communications model (with side information) can be used to reject the interference of $\bf x$ completely. Side-informed watermarking is typically instanced through quantization-based watermarking~\cite{wm:CW01}.    
In this case, the watermarked vector is obtained as 
\begin{equation}
\label{eq:qb}
    {\bf y} = Q_b({\bf x})
\end{equation}
where $Q_b(\cdot)$ is a secret (to adversaries) vector quantizer that depends on the embedded bit of information and is designed in such a way that ${\bf y}-{\bf x}$ meets a perceptibility constraint. The theoretical basis for side-informed watermarking was derived from Gel'fand and Pinsker's random-binning idea~\cite{wm:GP80} which relies on an auxiliary random variable as a proxy for trading off source and channel distortion. Especially influential to the IFS community was Costa's application of the random-binning paradigm to side-informed Gaussian channels that introduced the key element of distortion compensation in his construction of the auxiliary variable~\cite{wm:C83}. Following Costa's catchy title, the concept of coding with side information at the transmitter came to be known as ``dirty paper coding" (DPC). The rescue from the oblivion of DPC, now so prevalent in wireless communications, undeniably owes partly to watermarking research.

While the communications model serves as the core engine, the generic blueprint of a watermarking system typically requires additional components: a signal transformation to map the host media onto a multi-dimensional feature space that is robust to distortion, a powerful resynchronization framework to compensate for the misalignment experienced by the watermarked content, key-seeded pseudo-random mechanisms to obfuscate inner mechanics of the system to non-authorized parties, etc. This last item reveals a salient aspect of watermarking. A hostile adversary who wants to disrupt the watermark communications may be present, especially when watermarking is used for copyright protection. In that case, there is an incentive for pirates to strip the watermark that prevents them from accessing premium content or that somehow encodes their identity. 

Therefore, a sizable research effort has been dedicated to characterizing how to prevent such an adversary from detecting, estimating, tampering, and/or removing the watermark signal. For instance, it has been shown that specific measures should be taken to prevent watermark information leakage when the adversary has the opportunity to observe several watermarked assets~\cite{wm:CFT05}. On another front, research contributions have highlighted the risks of making the watermark detector publicly accessible. In that case, the adversary could devise powerful strategies to disrupt the watermark thanks to the availability of a reliable oracle~\cite{wm:CPP06}. This is particularly relevant when watermarking is used for copy or playback control, and relevant countermeasures must be implemented. More critically, findings from deployments for live video distribution revealed that pirate operators might blend different sources of the same video stream, each one with its watermark, to generate the video they distribute. Such adversarial behaviors, which have long been thought to be academic mind games, require the introduction of dedicated coding mechanisms for the watermark to survive, such as, for instance, anti-collusion codes~\cite{wm:TWW+03,wm:T08}.

Content protection applications' use cases have historically driven research on digital watermarking. For instance, in the late 90s, the Content Protection Technical Working Group (CPTWG) and the Secure Digital Music Initiative (SDMI) considered watermarking to implement a copy-control mechanism for DVDs and music. Still, the adoption of watermarking was hampered in its early days by controversies, e.g., the backlash against protection mechanisms after the US Digital Copyright Millennium Act or the EU Copyright Directive, and bullish marketing that oversold watermarking as a silver bullet against piracy. Nevertheless, forensic watermarking is widely deployed in digital cinemas, post-production houses, screeners systems, and direct-to-consumer video distribution platforms to trace the source of piracy. This commercial success has been further recognized in 2016 with a Technical Emmy Award for ``Steganographic Technologies for Audio/Video for Engineering Creativity in the Entertainment Industry." Digital watermarking is also routinely used to perform audience measurement for radio and TV by companies such as Nielsen in North America and M{\'e}diam{\'e}trie in Europe. 
Besides this core market, the scope of watermarking applications has now expanded beyond content protection, e.g., to convey metadata and media content reliably. For instance, Digimarc is pushing watermarking to replace barcodes in retail stores to speed up checkout time and is currently exploring if it could be used for plastic packaging to facilitate waste recycling.

\section{Robust Hashing}
\label{sec:robustHash}

The fundamental requirement of digital watermarking is that multimedia content needs to be modified prior to its delivery to the recipient. In other words, this technology degrades to some extent the content, which can appear odd in view of its goal to protect the asset. A parallel line of research in the IFS community, coined as robust hashing (also known as perceptual hashing or content fingerprinting), stemmed from this contradiction.

Robust hashing constructs a binary representation of the content in a robust low-dimensional space aiming at a fast and reliable recognition under severe distortions. It is common practice for watermarking techniques to also leverage such low-dimensional spaces to introduce the watermark signal in perceptually significant portions of the signal, and thereby achieve robustness. Without surprise, the bodies of work of both research areas therefore share several design patterns e.g. invariant spaces, pseudo-random projections, differential features, etc. On another front, there are connections to research on biometrics and content indexing that look for functions that output the same binary value for similar contents. Nevertheless, the IFS community undertook this challenge with a slightly different approach, inspired by cryptographic hash functions. These one-way functions have the property to produce very different hash values as soon as a single bit of the input changes. They are routinely used to check the integrity of digital assets and to provide efficient inverse lookup mechanisms for large scale databases. Robust hashing relaxes this `high sensitivity' property, and the aim is rather to tailor one-way functions that yield the same result for \emph{perceptually similar} pieces of content~\cite{fp:SMW06}. The underlying rationale is that a media asset should hash to the same or similar value even after modifications to the content that do not alter its semantics, such as recompression, filtering, resizing, and more. As such, robust hashing can be viewed as some quantization scheme in a robust multidimensional space.

A common approach, nowadays, is to have several such quantizers that produce sub-hashes combined with efficient nearest-neighbour search mechanisms. For instance, it has been exemplified that capturing the sign of the first derivative of some robust transform coefficient provides a rather stable hash representation of audio content~\cite{fp:HK02}. Such mechanism can then be exploited to construct song recognition applications, such as Shazam, or to provide means for broadcast monitoring or audience measurement. Robust hashing has also been used to check the integrity of multimedia documents or to identify physical object thanks to the hash of their micro-structures. Today, research on robust hashing mostly focuses on visual content, and is mostly published in content indexing and retrieval venues. A seminal example represents content as a bag of (visual) words~\cite{fp:SZ03}, each word being a robust hash, and exploits these words to recognize content. For instance, YouTube deploys such techniques to assess whether uploaded user generated content contains material subject to copyright claims.

\section{Steganography and Steganalysis}


Steganography is a tool for private covert communication. The sender typically hides a secret message in a host (cover) document by slightly modifying it and then communicating it overtly to the recipient. Steganographic channel is considered secure when an adversary observing the communication cannot detect the fact that steganography is being used. Once the use of steganography is detected, it is considered broken.

Steganography complements encryption for situations when even the existence of the private communication must be concealed and thus finds use in oppressive regimes that ban the use of encryption or for military operations. Statistical undetectability is the main factor distinguishing between steganography and watermarking. In contrast to steganography, the presence of watermarks is often advertised, and they usually have to be robust to distortion while carrying a relatively small payload.

Since detection of steganography amounts to detecting slight modifications of the host signal, steganalysis can be categorized as a forensic technique whose goal is to establish whether the host signal has been modified in a way that is indicative of embedding a secret. Consequently, many techniques developed for steganalysis found applications in forensics and vice versa.

\subsection{Steganographic Security}
\label{sec:Steg-security}

\begin{figure}
	\begin{center}
		\pgfplotsset{compat=newest, width=0.8\textwidth} 
		\begin{tikzpicture}
			\draw (0,0) node[draw, rectangle,text width=2.35cm,text centered] (emb) {$Emb(\mathbf{c}, \mathbf{m}, \mathbf{k})$};
			\draw (0,1) node[draw,rounded corners=5pt,black,text height=2.5mm] (msg) {message $\mathbf{m}$};
			\draw (0,-1) node[draw,rounded corners=5pt,black] (key1) {key $\mathbf{k}$};
			\draw (-2.3,0) node[draw,rounded corners=5pt,black] (cover) {cover $\mathbf{c}$};
			
			\draw (7.9,0) node[draw,text centered,minimum height=0.7cm] (extr) {$Ext(\mathbf{s}, \mathbf{k})$};
			\draw (7.9,-1) node[draw,rounded corners=5pt,black] (key2) {key $\mathbf{k}$};
			
			\draw[-latex] (msg)--(emb);
			\draw[-latex] (key1)--(emb);
			\draw[-latex] (cover)--(emb);
			
			\draw[-latex] (key2)--(extr);
			\draw[-latex] (extr)-- ++(0,0.8) node[above,draw,rounded corners=5pt,black,text height=2.5mm] {message $\mathbf{m}$};
			
			\draw [->,decorate,decoration={snake,amplitude=1.2mm,segment length=10mm},line width=2pt,black!50,-latex] (emb) -- (extr);
			\draw (emb) ++(4.0,0) node[below=0.4cm,text centered,text width=3cm] {\textnormal{channel with passive warden}};
			\draw (emb) ++(4.0,0) node[draw,rounded corners=8pt,line width=2pt,color=black!50,text=black,fill=white,text width=1.5cm,text centered] {stego $\mathbf{s}$};
			
			\draw (0,2.0) node[draw=none,font=\large] (Alice) {Alice};
			\draw (4.0,2.0) node[draw=none,font=\large] (Eve) {Warden};
			\draw (7.9,2.0) node[draw=none,font=\large] (Bob) {Bob};
			\draw[black!60, densely dashed] (2.1,-1.3)--(2.1,2.3);
			\draw[black!60, densely dashed] (5.9,-1.3)--(5.9,2.3);
			
		\end{tikzpicture}
	\end{center}
	
	\caption{\label{fig:steganographic_channel}Components of the steganographic channel.}
\end{figure}
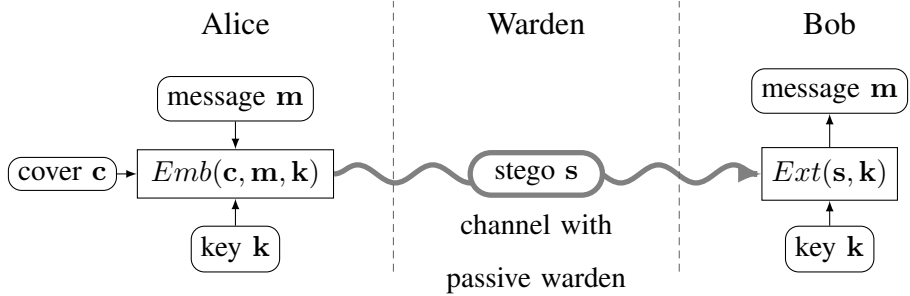

Steganography in its modern form is conceptually nested within the field of information theory, statistical hypothesis testing, and coding. The senders, usually named Alice and Bob, communicate by exchanging objects, which we will assume are digital images. They both agree upon the embedding and extraction algorithms used to embed a secret message in a cover and extract it from the stego object. Both algorithms make use of a secret shared key as depicted in Fig.~\ref{fig:steganographic_channel}.

Covers are drawn for communication according to some probability distribution $P_{c}$. If Alice uses steganography, her images $\mathbf{s}$ will in general follow a different stego distribution $P_{s}$. If Alice is able to make sure that $P_{c}=P_{s}$, the stegosystem is considered perfectly secure because it is impossible to distinguish between $P_{c}$ and $P_{s}$ by just observing the channel.

Perfect security is only achievable when Alice knows the cover distribution, in which case she can synthesize images from her secret message by running it through a cover source entropy decoder. Bob reads the message by compressing the image.

For digital media, though, the underlying statistical model is never perfectly known and all stego schemes in practice are thus imperfect. A useful measure of steganographic security is the Kullback–Leibler (KL) divergence:
\begin{equation}
D_{\mathrm{KL}}(P_{c}||P_{s})=\sum_{\mathbf{x}}P_{c}(\mathbf{x})\log\frac{P_{c}(\mathbf{x})}{P_{s}(\mathbf{x})}\label{eq:KL-divergence}
\end{equation}
because it  bounds the performance of the best detector that the Warden can build. For imperfect steganography, $D_{\mathrm{KL}}(P_{c}||P_{s})>0$, and with $n$ images being sent, Alice needs to scale her payload to be proportional to $\sqrt{n}$ to avoid being caught with near certainty by the Warden. This asymptotic result is known as the square root law of imperfect steganography~\cite{steg:Ker17ih}.

\subsection{Practical Steganography}
\label{sec:Practical-steganography}

There are two main image formats currently in use - raster formats, such as BMP, PNG, or TIFF, and the popular JPEG format. Steganographic methods for JPEG files modify the quantized DCT coefficients to embed the secret message.

Modern embedding methods are adaptive to content -- they take into account the detectability of embedding changes in different parts of the cover image. Intuitively, changes made to blue sky, water, out of focus parts of the image, and overexposed pixels will be more detectable than in textured areas, such as sand, grass, and trees. Alice can guide the embedding by assigning `costs' of changing each cover element and then requesting that the expected total cost of embedding (distortion) is minimal. Alternatively, she can adopt a statistical model and embed while minimizing statistical detectability usually simplified to the point that it can be expressed using the deflection coefficient.

Today, virtually all steganographic techniques for digital images use some form of the above two paradigms. The actual embedding is implemented using linear codes with the message $\mathbf{m}$ being communicated as the syndrome of the stego image represented with bits $\mathbf{y} = \bmod{(\mathbf{s},2)}$: $\mathbf{m} = \it{Ext}(\mathbf{s}) = \mathbf{Hy}$, where $\mathbf{H}$ is the code parity-check matrix. The parity-check matrices of the so-called syndrome-trellis codes~\cite{steg:Fil10stc} offer a clever blend of randomness for optimality in terms of the payload-distortion (detectability) trade off with enough structure to allow computationally efficient implementation using the Viterbi algorithm.

\begin{figure}
	\begin{centering}
		\begin{tikzpicture}
			\begin{axis}[
				ybar,
				ymin=0.0,ymax=0.50,xmin=0,xmax=9,
				height = 5cm, width = 8cm,
				ylabel = {Detection error $P_{\mathrm{E}}$},
				bar width = 0.5cm,
				xtick = {1,2,3,4,5,6,7,8},
				x tick label style={rotate=50,anchor=east},
				xticklabels = {\tiny{2002},\tiny{2009},\tiny{2010},\tiny{2011},\tiny{2014},\tiny{2016},\tiny{2017},\tiny{2018}},
				ytick = {0.1,0.2,0.3,0.4,0.5},
				ymajorgrids
				]
				\addplot [fill=black!50] coordinates {
					(1,0.4485) 
					(2,0.3875) 
					(3,0.3767) 
					(4,0.2834) 
					(5,0.2412) 
					(6,0.2338) 
					(7,0.1708) 
					(8,0.1290) 
				};
			\end{axis}
		\end{tikzpicture}
		\par\end{centering}
	\caption{\label{fig:HILL-detection-history}Detection error for stego algorithm
		HILL at 0.4 bits per pixel. The improvements during 2011-2016 are due to use of rich models while CNNs are responsible for the advancements during 2017-2018.}
\end{figure}
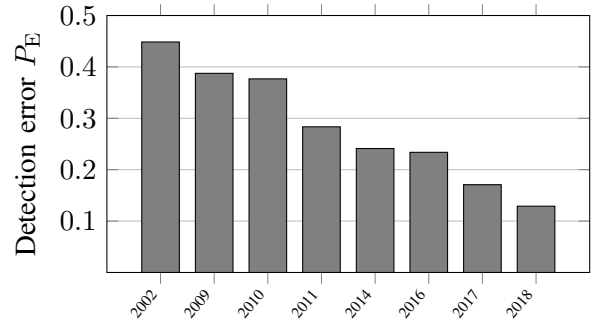

\subsection{Steganalysis}
\label{sec:Steganalysis}

Steganalysis detectors can be built either using the tools of detection theory as a form of the likelihood ratio test or they can be data-driven constructed using machine learning. The former usually imposes a statistical model on signals extracted from the image (typically noise residuals) while data-driven detectors are built by representing images using `features' hand designed to be sensitive to embedding changes but insensitive to image content.

A popular general methodology for designing feature representations for steganalysis is based on the concept of a rich model consisting of a large number of diverse submodels~\cite{steg:Fri11srm}. Rich media models can be viewed as a step towards automatizing steganalysis to facilitate fast development of accurate detectors of emerging steganographic schemes instead of having to develop a new approach for each new embedding method.

The latest generation of detectors are built in a purely data-driven fashion by presenting a deep Convolutional Neural Network (CNN) with examples of cover and stego images. This constitutes yet another paradigm shift in the field of steganalysis that lead to significant improvements in detection accuracy of just about every embedding scheme in both the spatial and JPEG domains (see an example of detection of HILL in Figure~\ref{fig:HILL-detection-history}).

The most recent trend in steganalysis (and in forensics in general) with deep learning is to use CNNs pretrained on computer vision tasks as a good starting point and apply the techniques of transfer learning to refine them
for steganalysis. For steganalysis, though, one needs to make sure that the feature map resolution is not decreased via pooling or strides too early in the network architecture as this form of averaging suppresses the signal of interest, the noise-like stego signal, while reinforcing the content, which is really the `corrupting noise' for the steganalyst~\cite{steg:yousfi20wifs}.

Deep learning has also advanced the field of steganography in the form of adversarial embedding and fully automatized data-driven learning of embedding costs BACKPACK~\cite{steg:BP21}.

\section{Biometrics}

In the last couple of decades, biometric technologies have become more and more pervasive in our everyday life due to several inherent advantages they offer over conventional recognition methods, which are based on what a person knows, \emph{e.g.} passwords or PINs, or what a person has, \emph{e.g.} ID cards or tokens. However, using biometric data raises many security issues specific to biometric-based recognition systems and not affecting conventional approaches for automatic people recognition. 

Biometrics such as voice, face, fingerprints, and iris, to cite a few, are exposed traits. This is because they are not secret; therefore, they can be covertly acquired or stolen by an attacker and eventually misused, leading, for example, to identity theft. Moreover, contrary to passwords or PINs, raw biometrics cannot be revoked, canceled, or reissued if compromised since they are the user's intrinsic traits and are limited in numbers. 

The use of biometrics also poses many privacy concerns. Biometric data can be used for purposes different than what they have been intended in the first instance of collection and for what the individual has agreed on. Moreover, when an individual gives out biometric characteristics, information about the person, like ethnicity, gender, and health conditions, can be potentially disclosed. Some biometrics can be covertly acquired at a distance and, therefore, could be used for surveillance. In addition, using biometrics as a universal identifier across different applications can allow person tracking, thus potentially leading to profiling and social control. To some extent, the loss of anonymity can be directly perceived by users as a loss of autonomy.

\begin{figure}
	\centering
	\includegraphics[width=1\linewidth]{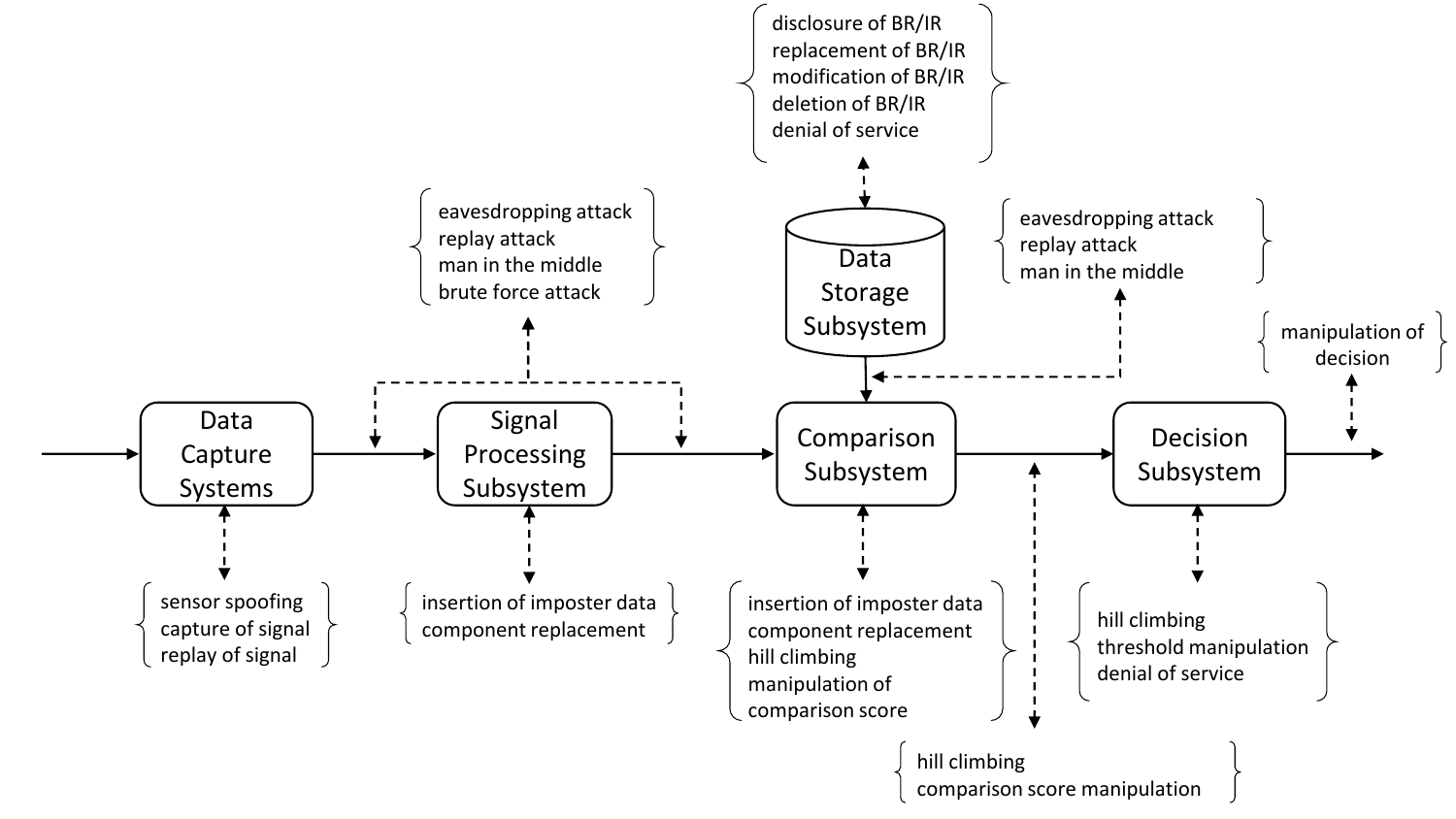} 
	\caption{Threats of a biometric system.}
	\label{fig:vulnerabilities}
\end{figure}

In recent years, the need to develop secure and privacy-compliant biometric systems has stimulated industrial and academic research~\cite{Cam13} and standardization activities~\cite{ISO24745}. A biometric system is the interconnection of the data capture, the signal processing, the comparison, the decision, and the data storage subsystems. Threats against a biometric system are diverse and can be perpetrated against its different components, including the transmission channels. In Figure~\ref{fig:vulnerabilities}, the major intentional attacks are synthetically illustrated.

Among the depicted attacks, those against the templates, hindering both the security and the privacy of biometric systems, and the spoofing attack, \emph{a.k.a.} presentation attack, at the sensor level, have become two mainstreams in biometric related research.

\subsection{Security and Privacy Requirements of a Biometric System}

The need to protect biometric templates has emerged as a very stringent requirement for the deployment of secure and privacy sympathetic biometric systems. However, classical encryption techniques cannot be effectively employed, essentially because of the noisy nature of biometric data that does not allow to make comparisons directly in the encrypted domain. To answer the need for secure and privacy compliant biometric systems, several approaches that treat the security and the privacy requirements as two facets of the same medal, aiming at enhancing security and minimizing privacy invasiveness, have been designed in the recent past also triggering a significant standardization activity.

The ISO/IEC 24745 Standard~\cite{ISO24745} makes a clear distinction between the \emph{identity reference} (IR) and the \emph{biometric reference} (BR), where the first refer to non-biometric attributes, such as name, address and so forth, uniquely identifying the user, and the second to biometric related attributes of the individual.

From an ideal standpoint, a secure and privacy-compliant biometric system needs to possess the following properties:
\begin{itemize}
    \item {\em{Confidentiality}}: from a security point of view, the biometric reference can be made available to authorized entities, who have full control of the data, and are protected from non-authorized ones. From a privacy perspective, the biometric reference can only be accessed by the entities who need to access the data and for the purpose they have been initially collected. The user has full control of the data with the right to be forgotten.
    \item {\em{Integrity}}: the integrity of the biometric sample, of the biometric reference, and the whole authentication process, need to be ensured.
    \item {\em{Revocability and Renewability}}: if a biometric reference is compromised, severe security and privacy issues can arise since an attacker can get unauthorized access, or the biometric reference or other personal data can be revealed. Therefore, it should be possible to revoke a compromised biometric reference and reissue a new one based on the same biometric sample.
    \item {\em{Irreversability}}: To prevent the biometric data from being used for purposes different from the originally intended and agreed ones, the biometric reference needs to be irreversibly transformed before being stored.
    \item {\em{Unlinkability}}: the biometric references should not be linkable, and it should not be possible to infer that they originated from the same biometric data, in a computationally feasible way, for different applications or datasets. 
\end{itemize}
In addition, from an ideal point of view, the performance of a biometric system compliant with the requirements above should not degrade. 

\subsection{Biometric Template Protection}

According to the ISO/IEC 24745 Standard, a renewable biometric reference consists of two elements, the \emph{pseudonymous identifier} (PI) and the \emph{auxiliary data} (AD), which are generated during the enrollment phase and kept separated either logically or physically. 

Biometric template protection schemes can be broadly classified into two different categories, specifically \emph{transformation-based} approaches and \emph{biometric cryptosystems}. It is worth mentioning that hybrid methods combining the above two have also been investigated.

Transformation-based approaches
rely on using a function whose parameters represent the AD, either invertible or non-invertible, used to transform the biometric reference. The original template is discarded, whereas the transformed one is stored as PI. The match is then performed in the transformed domain. The employed function can be invertible, in which case a key is needed, and the security of the approach depends on key management. Alternatively, systems relying on the use of non-invertible transformations are more secure, but the design of a non-invertible transformation not degrading the recognition performance accuracy is challenging. Several transformations have been proposed ever since and have been applied to a variety of biometric identifiers \cite{MCFGN10,PRC15}. 

Biometric cryptosystems are based on adapting cryptographic techniques to the intrinsic noisy nature of biometric data, usually employing error correction coding approaches to protect a biometric reference. Roughly speaking, biometric cryptosystems can be classified into \emph{key generation}, where a key is obtained from the biometric reference, and \emph{key binding} schemes, where the biometric reference is bound to a key. In both cases, a secret related to the PI is bound to the biometric reference to generate some public data, namely the AD, that ideally does not leak information about the biometric reference and that, in conjunction with the biometric reference, can allow retrieving the secret. Within the key binding scenario, the introduction of the two constructions, the \emph{fuzzy commitment}
for ordered data, and the \emph{fuzzy vault}
for unordered data, have significantly stimulated research in the field and have been applied to several biometric identifiers~\cite{NJ15}. In~\cite{DRS04}, an error-tolerant cryptographic primitive, namely the \emph{secure sketch}, has been introduced, and in~\cite{SLM07}, a general framework for analyzing the security of a secure sketch with application to face biometrics is carried out.

Approaches based on secure multiparty computation, employing homomorphic encryption and garbled circuits, have also been proposed for secure and privacy-compliant biometric systems~\cite{6461629}. These architectures rely on the computation of the distance between the stored biometric reference and the probe biometrics in the encrypted domain. The privacy and security of these approaches depend on the computational effort to discover a decryption key by an adversary. However, these methods have shown not to be mature jet to be deployed in real-life applications, like those requiring fast identification response, since the involved computational complexity, communication burden, and computational time are significantly higher than other architectures.

Several information-theoretic studies have investigated the potential AD information leakage in key binding approaches, that is, the amount of information leaked by the AD about the biometric reference.  
In~\cite{IW09}, 
the fundamental trade-off between the secret-key and privacy-leakage rates in biometric systems is studied for different scenarios. In~\cite{IW15}, the findings of \cite{IW09} are expanded by further discussing the trade-off among security, privacy, and identification performance. It has been pointed out that the larger identification rates are achieved, the more information leakage must be tolerated, and the smallest secret keys can be generated.
In addition, the need to provide a quantitative evaluation of the unlinkability has been addressed in ISO/IEC 30136~\cite{ISO30136}, where metrics for security and privacy protection performance assessment have been given.

\subsection{Presentation Attacks} 

Due to the integration of biometric sensors in almost every smart device, and their use in several applications, presentation attacks~\cite{MNFE19}, defined as the presentation of previously stolen human characteristics or fake ones to the acquisition sensor of a biometric system to gain unauthorized access, are receiving an increasing interest. Several approaches have been proposed mainly for fingerprint, iris, and face biometrics. The advent of deep learning has further fed this line of research also with the development of deepfakes (Section~\ref{deepfakes}).

\section{Multimedia Forensics}
\label{sec:forensics}

The analysis of multimedia evidence has been an essential part of digital forensics since the 80s. However, only in the late 90s, with the diffusion of personal digital devices, it became a full-fledged research field known as multimedia forensics, with a focus on source identification (for example, establishing which camera took a given photo) and authenticity verification (for example, detecting the presence and position of manipulated areas in an image). These areas have radically evolved in the last 25 years, following the equally fast evolution of key enabling technologies, such as the hardware and software of imaging devices or new methods for synthetic data generation, and pushed by the massive increase in the volume of audio-visual communications over the internet and social networks.

The progression of the IFS research from digital watermarking to multimedia forensics revolves around the use of `extrinsic' vs. `intrinsic' features~\cite{CKM+09, SWL13}. These notions were first coined by an interdisciplinary team at the Purdue University for electrophotographic printers~\cite{AMA+03}. The team examined the banding artifacts of printers and treated them as an “intrinsic” signature of the printer that can be identified by appropriate image analysis techniques; they also developed a way to manipulate the banding artifact to embed additional information as an “extrinsic” signature to encode side information such as the date and time that a document was printed.

\subsection{From Statistical to Data-driven Approaches}
Around the turn of the millennium, the most popular approaches for source identification and forgery detection relied on mathematical or statistical models. 
A breakthrough in the field was the emergence of methods based on the concept of device fingerprint, following the seminal 2006 work~\cite{LFG06} on the camera photo response nonuniformity (PRNU) pattern. The PRNU is a deterministic sensor pattern due to tiny imperfections in the sensor manufacturing
and can be regarded as a sort of device fingerprint. PRNU-based methods significantly advanced the state-of-the-art in both source identification and image forgery detection
and have been extensively used by law enforcement agencies 
to analyze both physical devices and web accounts.
Many other pieces of information have been exploited for forensic investigation, arising from all phases of the digital life of multimedia assets: traces of in-camera operations, such as chromatic aberrations, color filter array artifacts, and double JPEG compression, 
but also clues related to out-camera image processing steps, like image smearing, shadows and reflections~\cite{Far09}. 

An inherent limit of model-based approaches 
is that they mostly fail when the hypotheses do not hold. On the other hand, this is rather the norm in real-world uncontrolled scenarios where the data go through unpredictable post-processing operations, as it happens on social networks.
Another major issue is technological advances. For example, the introduction of computational photography has strongly changed the data acquisition pipeline and
many hypotheses of model-based methods do not hold anymore.
Data-driven machine learning methods can partially solve these problems, and in fact, they have been successfully applied 
starting from the first experiments in 2005~\cite{SWL13}. 
Several tools also took inspiration from work carried out in steganalysis since, despite the obvious differences, both research areas focus on seemingly invisible alterations of the natural characteristics of an image. 
For example, in the 2013 IEEE Image Forensics Challenge, the winning solutions relied on the rich models~\cite{steg:Fri11srm} developed with great success in steganalysis.

\subsection{The Advent of Deepfakes}
\label{deepfakes}  
Deep learning has brought a revolution in multimedia forensics, making available a wealth of simple and powerful tools that allow one to create synthetic content easily. 
The first deepfake video dates back only to autumn 2017. However, the widespread diffusion of tools based on autoencoders, generative adversarial networks, and, more recently, diffusion models has led to the exponential growth of deepfakes we observe today, which threatens so many fields of our society, from politics to journalism to the private lives of citizens. In particular, AI-powered tools that allow one to generate realistic faces of persons 
has raised great alarm not only for the diffusion of misinformation but also for the vulnerability of biometric systems.

However, deep learning impacted heavily also on the defense side, making new powerful tools and methodologies available to the forensic analyst. Especially important was the creation of larger and larger datasets of manipulated media, which allowed researchers to train or fine-tune deep neural networks~\cite{Ver20}. Deep learning-based detectors, trained and tested on such datasets, soon outperformed methods that relied on handcrafted features. In particular, in the most challenging situations of low-quality compressed videos there is a large gap between a solution based on extracting forensic features and a fully data-driven method based on a very deep convolutional neural network~\cite{RCV+19}. In attribution tasks, deep learning allowed one to learn a camera fingerprint from the available data, gaining independence from a fixed mathematical model and proving more effective in many different situations~\cite{CL20}. The concept of fingerprint was also extended to synthetic images, with the so-called artificial fingerprints, related to new types of artifacts introduced in the generation process~\cite{Ver20}.

Despite their obvious potential, AI-based methods also have well-known weaknesses, from a general lack of interpretability to a limited generalization ability, with poor performance on data generated by manipulation methods or synthetic sources never seen in the training phase. A further major challenge is represented by adversarial attacks, which can easily fool deep learning detectors. 
This happens especially when the detector relies on low-level features that can be easily removed by injecting suitable adversarial noise. 
For this reason, a recent trend is towards the exploitation of 
semantic artifacts, more robust to different signal degradations, such as the biometrics of a specific identity or the geographic information estimated from an image or video. 

\subsection{From Single to Multimodal Analysis}

Another major evolution is represented by a paradigm shift from data processed in isolation to multimodal analyses. 
With the progress of technology, devices tend to lack unique features that allow easy identification, and manipulations become increasingly sophisticated, evading the scrutiny of expert users. In this context, working on a single media modality may be insufficient, while 
a joint analysis of all pieces of information associated with a media asset may become key to successful forensics. Accordingly, current methods look for inconsistencies between multiple modalities, such as audio-video or text-image, the latter being especially relevant when unaltered images are used in a new but false context. This trend started already in the early '10s, with the introduction of multimedia phylogeny \cite{DRG11}, which 
aims at investigating the history and evolutionary process of digital objects by automatically identifying the structure of relationships underlying the data.
This has led to a synergy between different research fields: signal processing, computer vision, and natural language processing. In parallel, the attention to multimedia forensics has moved from forensics labs and law enforcement agencies, as it was 25 years ago, 
to big tech companies, such as Facebook and Google, and large international research programs. 
Likewise, research papers once published mostly in specialized forensics venues now find a much wider audience, including major computer vision conferences and satellite workshops.
\section{Adversarial Signal Processing and Machine Learning}

If there is one thing that researchers trained in the watermarking gym had learned by the end of the first decade of the new millennium, it is that security is not robustness~\cite{PCT+06}. Dealing with random noise and benevolent manipulations is not like dealing with an enemy whose explicit goal is to make the system fail.
In the meantime, other security-oriented signal-processing applications were emerging, including multimedia forensics \cite{BST18}, biometric security \cite{BRD+15}, network intrusion detection \cite{MMT08}, spam filtering, anomaly detection, and many others. Despite their differences, all these fields were characterized by a unifying feature: the possible presence of one or more adversaries aiming at making the system fail.
Prompted by this basic observation, multimedia security researchers started studying the adversarial dynamics describing the interplay between the actions of the system designer and the adversary. Some early works in this area include \cite{ZL06,ZLL11}, where game theory was used to predict behavioural dynamics in traitor tracing and media sharing applications.
As a result of these activities, a broad research area, often referred to as {\em adversarial signal processing}~\cite{BP13}, emerged, whose final goal is to design signal processing tools that retain their effectiveness even in the presence of an adversary.

The peculiar feature of adversarial signal processing is the presence of an {\em informed} and {\em intelligent} attacker, that does not act stochastically, since he/she introduces a disturbance optimized to cause the maximum damage to the system. To do so, the attacker exploits the knowledge he/she has about the to-be-attacked system.
As a leading example (and without pretending to be exhaustive), let us consider a system responsible for making a binary decision.
The binary decision may regard the presence of a watermark within a signal, detecting anomalous behavior, or verifying a biometric trait.
Let us assume, for simplicity, that a linear function is used. More specifically, let
\begin{equation}
    \phi({\bf x}) = \left< {\bf x}, {\bf w} \right> = \sum_{i=i}^n x(i) w(i),
\label{eq.lindet}
\end{equation}
be a linear combination of the input signal $\bf{x}$ and a proper vector of weights $\bf{w}$. The system makes a positive decision if $\phi({\bf x}) > T$, and a negative one otherwise. In digital watermarking, for instance, ${\bf x}$ corresponds to the observed signal and ${\bf w}$ to the watermarking key.
As another example, eq. \eqref{eq.lindet} may model an anomaly detector based on Fisher discriminant analysis. In this case, ${\bf x}$ contains the features the detector relies on and ${\bf w}$ the weights of the linear combination.
In a white-box attack, the attacker knows exactly the form of $\phi$, including the vector ${\bf w}$. In this case, the optimum attack corresponds to adding a perturbation ${\bf z}$ defined as follows:
\begin{equation}
    {\bf z} = (T - \left< {\bf x}, {\bf w} \right> - \varepsilon) \frac{{\bf e}_w}{\| {\bf w} \|},
\label{eq.optlinattack}
\end{equation}
where ${\bf e}_w$ is a vector having the direction of ${\bf w}$, $\varepsilon$ is an arbitrarily small positive quantity, and where we have assumed that the goal of the attacker is to turn a positive decision into a negative one.
If the attacker does not know ${\bf w}$, attacking the system would require the addition of random noise with two negative consequences (for the attacker): i) uncertainty about the result of the attack, and ii) necessity of introducing into the system a larger distortion (on the average).

Alternatively, the attacker may not know ${\bf w}$, but may have access to the values assumed by $\phi$ in correspondence with properly chosen inputs. In this case, the attacker may estimate the gradient of $\phi$ and add a perturbation aligned to the negative direction of the gradient. If $\phi$ is nonlinear, he/she can use gradient descent to exit the positive decision region with a minimal distortion.
In other cases (black box attacks), the attacker can only observe the final decision of the system in correspondence to chosen inputs. In this case, he/she can apply a so-called sensitivity attack. Let us assume again that $\phi$ is a linear function. The attacker first chooses a random input resulting in a negative decision. Then, he/she applies a bisection algorithm to find a point on the boundary of the decision region. Finally, he/she repeats the procedure $n$ times, finding $n$ points on the boundary of the decision region. Due to the linearity of $\phi$, such $n$ points are enough to estimate ${\bf w}$ and compute ${\bf z}$ as in~\eqref{eq.optlinattack}. If the detection boundary is not linear, then the attack is more difficult, yet still possible, as shown in~\cite{wm:CPP06}.

Let us now turn our attention to the defender. First, the defender may want to keep part of the system secret. By following a common practice in cryptography, secrecy should be incorporated within a key, while the overall form of the system is assumed to be known. In the simple linear case outlined before, this means that the attacker knows the form of $\phi$, but does not know ${\bf w}$. If the defender knows the attacker's strategy, he/she can adopt other countermeasures. For instance, he/she may try to limit the information the attacker can infer by observing the system's output by randomizing the decision function. In systems based on machine learning, the defender may retrain the system by incorporating some attacked inputs in the training set. In this way, the system learns to recognize the attacked inputs and treat them properly.

A problem with most of the defenses described so far is that they assume a static situation where the attacker adopts a fixed strategy ignoring the possible countermeasures adopted by the defender\footnote{No need saying that a similar drawback applies to most attacking strategy.}. 
When this is not the case, the defender can adopt a worst-case solution, assuming that the attacker has perfect knowledge of the attacked system. However, this tends to be an overly pessimistic approach, given that in some cases it may be difficult, or even impossible, for the attacker to obtain a perfect knowledge of the system. Furthermore, the defenses put forward under the worst-case assumption may be too complicated, leading to a significant deterioration of the system's performance. An elegant solution to solve this apparent deadlock and avoid that new attacks and defenses are developed iteratively in a never-ending loop, consists in modeling the interplay between the attacker and the defender using game theory.
Game theory, in fact, provides a powerful way to model the interplay between the attacker and the defender, whose contrasting can be defined by the payoff of a zero-sum game, while the constraints they are subject to and the knowledge they have can be modeled by a proper definition of the set of moves they can choose from. Furthermore, it is possible to model both scenarios wherein the attacker and the defender design their systems independently, and situations where one of the players acts first and the other adapts his/her move based on the choice made by the first player. Eventually, by computing the payoff at the equilibrium, the achievable performance of the system when both players adopt an {\em optimum} strategy can be evaluated.
Some examples of works where game theory was successfully used to derive the optimal strategies for the attacker and the defender include \cite{ZL06, SB12, MMT08, BT20}. 

\subsection{Adversarial AI}

Deep Learning (DL) and Artificial Intelligence (AI) are revolutionizing the way signals and data are processed. In addition to new opportunities, DL raises new security concerns. When Szegedy et al.~\cite{SZS+13} pointed out the existence of adversarial examples affecting virtually any deep neural network model, the AI community realized that robustness is not security and proper countermeasures had to be taken if AI had to be used within an adversarial setting. Such concerns gave birth to a new discipline, usually referred to as adversarial AI (or adversarial machine learning).
Adversarial machine learning has many similarities with adversarial signal processing. When targeting a binary classification network, for instance, adversarial attacks are nothing but a re-edition of the white-box attacks described in eq. \eqref{eq.optlinattack}. More generally, by assuming that the to-be-attacked input is close to the decision boundary and that the boundary is locally flat, in its simplest form, an adversarial example can be computed as:
\begin{equation}
    {\bf x}_{adv} = {\bf x} - \varepsilon {\bf e}_{\nabla{\phi({\bf x})}}, 
    \label{eq.advatt}
\end{equation}
where we have assumed that the goal of the attack is to change the sign of the network output, ${\bf e}_{\nabla{\phi({\bf x})}}$ is a vector indicating the direction of the gradient of the output of the network in correspondence of the input {\bf x}, and $\varepsilon$ is a (usually small) quantity ensuring that the network decision is inverted. Even if more sophisticated ways of constructing adversarial examples have been proposed, the similarity to adversarial signal processing is evident, the main peculiarity of attacks against DL architecture being the ease with which $\nabla{\phi({\bf x})}$ can be computed by relying on back-propagation.
A distinguishing feature of adversarial machine learning is the possibility of attacking the system during the learning phase. Backdoor attacks are a typical example where the attacker manipulates the training set to inject into the network a malevolent behavior~\cite{GTB22}. Interestingly, this attack presents several similarities with watermarking. Backdoor injection, in fact, can be seen as a way to watermark a neural network so to protect the IPR of DL models~\cite{LWB21}. 

Given the striking similarities between adversarial machine learning and security-oriented applications of signal processing, the signal processing community is in an ideal position to contribute to the emerging area of adversarial AI, transferring to this domain the theoretical and practical knowledge developed in the last 25 years.
\section{Additional Topics}

As mentioned earlier in the introduction, we opted for a focused review of the IFS research areas that experienced the most notable breakthroughs and that were aligned with the typical interest of the readership of the Signal Processing Magazine. Nevertheless, the scope of IFS is much wider and this section is intended to provide a glance at other relevant sub-areas. The most interested readers are invited to check out our flagship journal publication, the IEEE Transactions on Information Forensics and Security, to grasp a comprehensive view of the IFS domain including areas closer to Computer Science, Information Theory, and Digital Communications.

Due to its security-oriented application domain, a sizable portion of the IFS research has been related to applied cryptography for (multimedia) signals. While the IFS community has not been so much involved with advances for Digital Rights Management (DRM) and Conditional Access Systems (CAS), it explored alternate encryption mechanisms that would be better suited for multimedia signal compared to bulk encryption. This techniques, routinely coined as \emph{selective encryption}~\cite{misc:MLD+08} consists in encrypting only a small portion of the signal representation to incur unrecoverable degradation while preserving lossy compression capabilities. Unfortunately, the parsing cost of these techniques hampered their adoption. Surviving incarnations of this paradigm today comprise pattern-based, most notably used in Apple's Sample AES and MPEG Common Encryption~\cite{ISO23001}, and encryption limited to a region of interest.

On another front, signal processing in the encrypted domain received increasing attention in the late 2000s. The necessity of processing encrypted signals without first decrypting them arises naturally whenever two or more parties need to cooperate to reach a common goal, without revealing proprietary signals (and data) of a private nature, like, for instance medical records~\cite{LEB12}. The cryptography community provided baseline tools, namely homomorphic encryption and multiparty computations, to process encrypted data. Nevertheless, when the data to be processed take the form of signals, it is necessary to exploit synergies between cryptographic tools and signal processing techniques in order to obtain secure and efficient solutions. Over the years, the IFS community has greatly contributed to develop such solutions for a wide variety of applications domains including biomedical signal processing, biometrics, smart metering, private recommender systems, and many others.

While forensic techniques for multimedia documents have been discuss in Section~\ref{sec:forensics}, the IFS community has also explored how to apply similar methodology for other types of signals. For instance, human actions may lead to changes in the surrounding electromagnetic field associated with WiFi systems. These changes can then be analyzed to detect a variety of movements, e.g. (i) walking in a building or entering a room, (ii) subtle breathing movements, and (iii) unexpected sudden movement such as falling down~\cite{WXC+18}. The power grid, whose nominal frequency is varying over time in a unique manner, also provides a ubiquitous form of ambient signatures comes from the power grid. Harnessing the time-frequency properties of the grid signatures, that may be revealed in subtle but detectable ways from audiovisual recordings, can enable forensic analysis to determine the capturing time, location, and integrity of these recordings~\cite{GVH+13}.

A final line of research worth mentioning is the so-called \emph{security of noisy data}~\cite{TSK10}. It relates to the ability of reliably recognizing data and signals when their representation slightly differs from one observation to the other. This fundamental problem underpins several IFS sub-areas such as biometrics, robust hashing, and sensor-based forensics. Interestingly, the baseline tools developed to tackle this challenge can be revisited to tailor some kind of physically unclonable features (PUF). For instance, some IFS contributions have demonstrated that it is feasible to extract some signature from the microscopic structures of paper and other physical objects to facilite brand protection and the fight against counterfeiting~\cite{misc:DVH+14}.

\section{Conclusions}

All in all, within a quarter century, the IFS research community has widen its focus well beyond its initial goal that was revolving, for a large part, around intellectual property protection. It fully embraced the transition into our new digital world and addresses fundamental societal challenges related to trust, privacy, and protection. While such topics are routinely considered to be owned by computer science, the unique contributions of the IFS community clearly established that signal processing has its own role to play. For instance, the great strides in machine learning are anticipated to raise challenges with the emergence and popularization of generative methods capable of substantiating synthetic realities such as deepfakes and artificially-generated images, text, and videos. This blur of the frontier between natural and synthetic signals is happening right now and will undeniably become an exciting playground for the IFS research community.

While the changes coming with machine learning may be scary, they are also likely to come with their own batch of benefits for the IFS research area, which typically aims at isolating/detecting low-power signals whose characteristics may not be known beforehand. Similarly to other domains, technological advances need to be accompanied, in some occasions, by an evolution of the legal framework that governs our lives to diffuse the risk of unmanaged technical advances, including to rule the use of IFS technologies and avoid their misuse. The road to hell is paved with good intentions and a surveillance technology to protect the safety of people can be abused to invade privacy. It is of equal importance to educate citizens and raise their awareness to some of the dangers inherent to our new digital world without scaring them. As Descartes used to say, human senses can be misleading and one should not take at face value what he sees, read, or hear (on the Internet).

\section{Acknowledgements}

We gratefully acknowledge the support of this research by the Defense Advanced Research Projects Agency (DARPA) under agreement number FA8750-20-2-1004.
The U.S. Government is authorized to reproduce and distribute reprints for Governmental purposes notwithstanding any copyright notation thereon.
The views and conclusions contained herein are those of the authors and should not be interpreted as necessarily representing the official policies or endorsements, either expressed or implied, of DARPA or the U.S. Government.
This work has also received funding by the European Union under the Horizon Europe vera.ai project, Grant Agreement number 101070093, and is supported by a TUM-IAS Hans Fischer Senior Fellowship and by the PREMIER project funded by the Italian Ministry of Education, University, and Research within the PRIN 2017 program. A. Rocha also thanks the financial support of the São Paulo Research Foundation (Fapesp) through the D\'ej\`aVu project. F. P\'erez-Gonz\'alez's research is partially funded by the Spanish Ministry for Science and Innovation and NextGenerationEU/PRTR through project FELDSPAR (MCIN/AEI/10.13039/501100011033), and by Xunta de Galicia and ERDF under project ED431C 2021/47.

\bibliographystyle{IEEEtran}
\bibliography{IEEEabrv,ref}

\end{document}